\documentclass[pdftex,twocolumn,epjc3]{svjour3}

\usepackage{amsmath,amsfonts,amssymb}
\usepackage[utf8]{inputenc}
\usepackage{graphicx}
\usepackage{epsfig}
\usepackage{subfigure}
\usepackage{color}
\usepackage{comment}
\usepackage{slashed}
\usepackage{yfonts}

\newcommand{\be}{\begin{equation}} 
\newcommand{\ee}{\end{equation}}
\newcommand{\bea}{\begin{eqnarray}} 
\newcommand{\eea}{\end{eqnarray}}

\newcommand{\td}{{\rm d}}

\newcommand{\aend}{a_\text{end}}
\newcommand{\acrit}{a_\text{crit}}

\newcommand{\Tend}{T_\text{end}}

\journalname{Eur. Phys. J. C}

\begin{document}

\title{Scalar singlet dark matter in non-standard cosmologies}

\author{Nicol\'{a}s Bernal\thanksref{e1,addr1} \and Catarina Cosme\thanksref{e2,addr2} \and Tommi Tenkanen\thanksref{e3,addr3} \and Ville Vaskonen\thanksref{e4,addr4}}

\institute{Centro de Investigaciones, Universidad Antonio Nari\~{n}o, Carrera 3 Este \# 47A-15, Bogot\'{a}, Colombia\label{addr1} \and
Departamento de F\'{\i}sica e Astronomia, Faculdade de Ci\^encias da Universidade do Porto and\\Centro de F\'{\i}sica do Porto, Rua do Campo Alegre 687, 4169-007, Porto, Portugal\label{addr2} \and
Astronomy Unit, Queen Mary University of London, Mile End Road, London, E1 4NS, United Kingdom\label{addr3} \and
NICPB, R\"avala pst.~10, 10143 Tallinn, Estonia\label{addr4}}

\thankstext{e1}{nicolas.bernal@uan.edu.co}
\thankstext{e2}{catarinacosme@fc.up.pt}
\thankstext{e3}{t.tenkanen@qmul.ac.uk}
\thankstext{e4}{ville.vaskonen@kbfi.ee}

\date{Received: date / Revised version: date}

\maketitle

\begin{abstract}
We study production of dark matter (DM) in models with a non-standard expansion history. We consider both freeze-out and freeze-in mechanisms for producing the observed DM abundance in a model where the DM consists of scalar singlet particles coupled to the Standard Model sector via the Higgs portal. We show that a non-standard expansion phase can lead to a significant change in the DM abundance and therefore to observational ramifications. For example, for DM freeze-in the required portal coupling can be much larger, whereas for DM freeze-out much smaller values become allowed. We evaluate the relevant constraints and discuss prospects for direct detection of such DM.
\end{abstract}

\section{Introduction}

For a very long time, Weakly Interacting Massive Particles (WIMPs) have been among the best-motivated dark matter (DM) candidates. However, given that there are no observational hints of particle DM and only increasingly strong constraints on WIMP DM~\cite{Arcadi:2017kky}, it is natural to question the existence of WIMPs and start considering other options for the production and properties of DM.

A simple alternative to the standard WIMP paradigm is provided by relaxing the usual assumption that DM is a thermal relic, produced by the freeze-out mechanism in the early Universe. Assuming instead that DM particles never entered into thermal equilibrium with the Standard Model (SM) plasma, the present DM abundance may have been produced by the so-called freeze-in mechanism~\cite{McDonald:2001vt,Hall:2009bx,Bernal:2017kxu}, where the observed relic abundance results from decays and annihilations of SM particles into DM. Because of the feeble interaction strength that the mechanism requires, this kind of DM candidates are usually called Feebly Interacting Massive Particles (FIMPs).

Another simple way to evade the experimental constraints on DM is to consider non-standard cosmological histories, for example scenarios where the Universe was effectively matter-dominated  at an early stage, due for example to slow reheating period after inflation or to a massive metastable particles. As there are no indispensable reasons to assume that the Universe was radiation-dominated  prior to Big Bang Nucleosynthesis (BBN)\footnote{For studies on baryogenesis with a low reheating temperature or during an early matter-dominated  phase, see Refs.~\cite{Davidson:2000dw,Giudice:2000ex,Allahverdi:2010im,Beniwal:2017eik,Allahverdi:2017edd} and~\cite{Bernal:2017zvx}, respectively.} at $t\sim 1$~s, studying what consequences such a non-standard era can have on observational properties of DM is worthwhile. Indeed, production of DM in scenarios with a non-standard expansion phase has recently gained increasing interest, see e.g. Refs.~\cite{Co:2015pka,Berlin:2016vnh,Tenkanen:2016jic,Dror:2016rxc,Berlin:2016gtr,DEramo:2017gpl,Hamdan:2017psw,Visinelli:2017qga,Drees:2017iod,Dror:2017gjq,DEramo:2017ecx,Bernal:2018ins,Hardy:2018bph,Maity:2018exj,Hambye:2018qjv}. For earlier works, see Refs. \cite{Kamionkowski:1990ni,Salati:2002md,Comelli:2003cv,Rosati:2003yw,Gelmini:2006pw,Gelmini:2006pq,Arbey:2008kv,Arbey:2009gt,Arbey:2018uho}.

In this paper, we will consider production of DM in scenarios where 
for some period at early times the expansion of the Universe was governed by a fluid component with an effective equation of state $p=w\,\rho$, where $p$ is the pressure and $\rho$ the energy density of the fluid, and $w\in [-1,1]$. For generality, we will consider production of DM by both the freeze-out and freeze-in mechanisms. Therefore, we have two goals: shed light on production of DM during a non-standard expansion phase in general, and study in detail the observational and experimental ramifications such a phase can have on the parameter space of a model where the DM consists of real singlet scalar particles $S$ coupled to the SM sector via the Higgs portal interaction $\lambda_{\rm HS} S^2 |H|^2/2$, where $H$ is the SM Higgs doublet and $\lambda_{\rm HS}$ a dimensionless coupling constant. We will then contrast our results with the earlier studies on the production of singlet scalar DM in the case of standard radiation-dominated cosmological history~\cite{Cline:2013gha,No:2013wsa,Alanne:2014bra,Robens:2015gla,Feng:2014vea,Han:2015hda,Athron:2017kgt}. We will also consider prospects for detection of such non-standard DM, including collider and direct detection experiments -- where the DM candidate is the usual thermal relic or has a non-thermal origin.

Recent studies in Refs.~\cite{Bernal:2018ins,Hardy:2018bph}, have shown that in this simple framework one can both evade the current observational constraints but expect to detect a signal in the near future. However, in order to fully understand the scenario and its observational prospects, a more detailed analysis than what was conducted in Refs.~\cite{Bernal:2018ins,Hardy:2018bph} is needed. In this paper, we therefore conduct a numerical study, considering a broad range of DM masses and sub-leading corrections to the cross-sections and decay rates relevant for the singlet scalar model, as well as taking into account the evolution of 
the effective number of SM energy density degrees of freedom. In contrast to the earlier studies, which concluded that even in the case where DM was produced by the freeze-in mechanism it may be possible to observe it by the means of direct detection, our results indicate that the parameter space relevant for freeze-in in the singlet scalar model is even in very extreme scenarios out of reach of the future direct detection experiments.

The paper is organized as follows: In Section~\ref{sec:model} we introduce the cosmological setup and the singlet scalar DM model, discussing various constraints on the parameter space of the model. Then, in Section~\ref{sec:DMproduction}, we conduct our numerical analysis for DM production in different cases, discussing also the effects of non-vanishing DM self-interactions, and contrast our results with the standard radiation-do\-mi\-na\-ted case. Finally, we present our conclusions in Section~\ref{sec:conclusions}.

\section{The model and constraints}
\label{sec:model}

\subsection{Expansion history}
\label{expansion}

We assume that for some period of the early Universe, the total energy density was dominated by a component $\rho_\phi$ with an equation of state parameter $w\in [-1,1]$, where $w\equiv p_\phi/\rho_\phi$, with $p_\phi$ the pressure of the dominant component. We assume that this component decays solely into SM radiation with a rate $\Gamma_\phi$ that, in general, is a function of time. Moreover, we assume that the SM plasma maintains internal equilibrium at all times in the early Universe.

In the early Universe the contribution of the DM energy density can be neglected, so the evolution of the energy densities $\rho_\phi$ and $\rho_{\rm R}$ are governed by the system of coupled Boltzmann equations
\be\label{BE0}
\begin{aligned}	
	\frac{\td\rho_\phi}{\td t} + 3(1+w)H\rho_\phi =& -\Gamma_\phi\,\rho_\phi \,, \\
	\frac{\td\rho_{\rm R}}{\td t} + 4H\rho_{\rm R} =& +\Gamma_\phi\,\rho_\phi \,,  
\end{aligned}
\ee
where $\rho_{\rm R}$ is the SM energy density. The Hubble expansion rate $H$ is defined by
\be\label{Hubble}
H^2=\frac{\rho_\phi+\rho_{\rm R}}{3\,M_P^2} \,,
\ee
where $M_P$ is the reduced Planck mass. Under the assumption that the SM plasma maintains internal equilibrium, the time (or scale factor) dependence of its temperature can be obtained from 
\be \label{rhoRT}
\rho_{\rm R} = \frac{\pi^2}{30} g_*(T)\, T^4\,.
\ee
Here $g_*(T)$ corresponds to the effective number of SM energy density degrees of freedom, which we evaluate as given in Ref.~\cite{Drees:2015exa}.

Consider first a constant $\Gamma_\phi$. This describes usual particle decay and approximates well, in some cases, also the decay of a time-evolving background field, such as an inflaton field during a reheating phase~\cite{Linde:2005ht}. The SM energy density evolves as a function of the scale factor $a$ as
\be\label{rhoRconstG}
\frac{\rho_{\rm R}(a)}{\rho_{\rm R}(\aend)} \simeq 
\begin{cases}
\left(\frac{a}{\aend}\right)^{-4} F^{-1} \,,  &a < \acrit \,, \\
\left(\frac{a}{a_{\mathrm{end}}}\right)^{-\frac32(w+1)} \,,  &\acrit \leq a <  \aend \,, \\
\left(\frac{a}{a_{\mathrm{end}}}\right)^{-4} \,,  &\aend \leq a \,,
\end{cases}
\ee
where $F$ describes how much the co-moving SM radiation energy density increases by decay of $\rho_\phi$, that is 
\be\label{eq:F}
F\equiv\frac{a^4\,\rho_{\rm R}(a)\big|_{a\gg \aend}}{a^4\,\rho_{\rm R}(a)\big|_{a\ll \acrit}}\,.
\ee 
At $a = \acrit$ the production from $\rho_\phi$ starts to dominate the evolution of $\rho_{\rm R}$, and at $a = \aend$ the $\rho_\phi$ dominated phase ends. The scaling of $\rho_{\rm R}$ for $a_{\rm crit}<a<a_{\rm end}$ follows from the Boltzmann equation for $\rho_{\rm R}$
\be
\frac{H}{a^3} \frac{\td}{\td a} \left(a^4 \,\rho_{\rm R}\right) = \Gamma_\phi\,\rho_\phi \,,
\ee
where for $a<a_{\rm end}$ the Hubble rate scales as $H\sim\sqrt{\rho_\phi}$ and $\rho_\phi\sim a^{-3(w+1)}$. 

The temperature of the SM plasma at $a=\aend$ is given by the total decay width $\Gamma_\phi$ as
\be
\Tend^4 = \frac{90 M_P^2\, \Gamma_\phi^2}{\pi^2\,g_*(\Tend)} \,.
\ee
For having successful BBN, the temperature at the end of the $\rho_\phi$ dominated phase has to satisfy $T_{\rm end} > 4$~MeV \cite{Kawasaki:2000en,Hannestad:2004px,Ichikawa:2005vw,DeBernardis:2008zz}.

\begin{figure*}[t]
\begin{center} 
\includegraphics[height=0.32\textwidth]{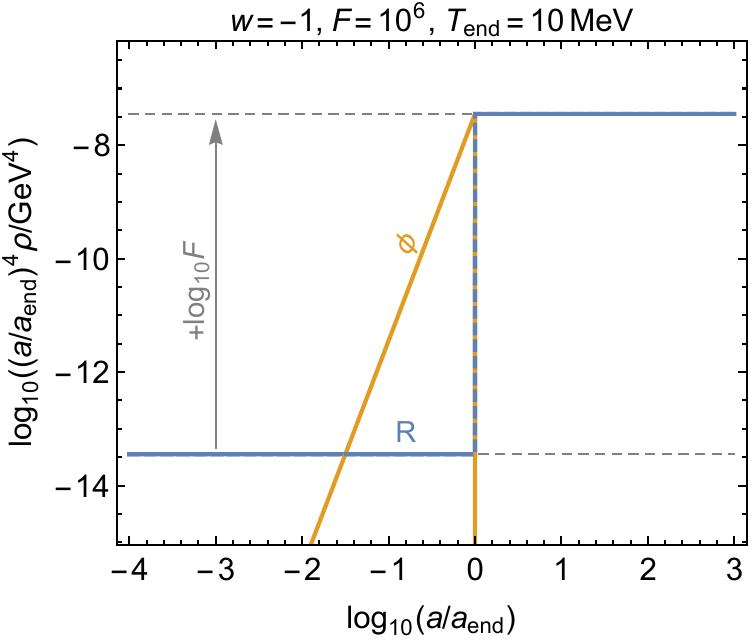} \hspace{6mm}
\includegraphics[height=0.32\textwidth]{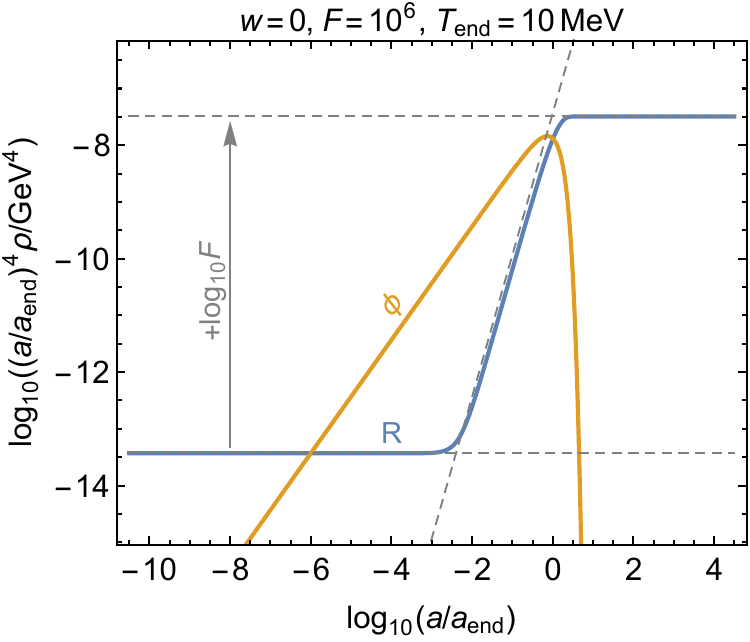} \\ \vspace{2mm}
\includegraphics[height=0.32\textwidth]{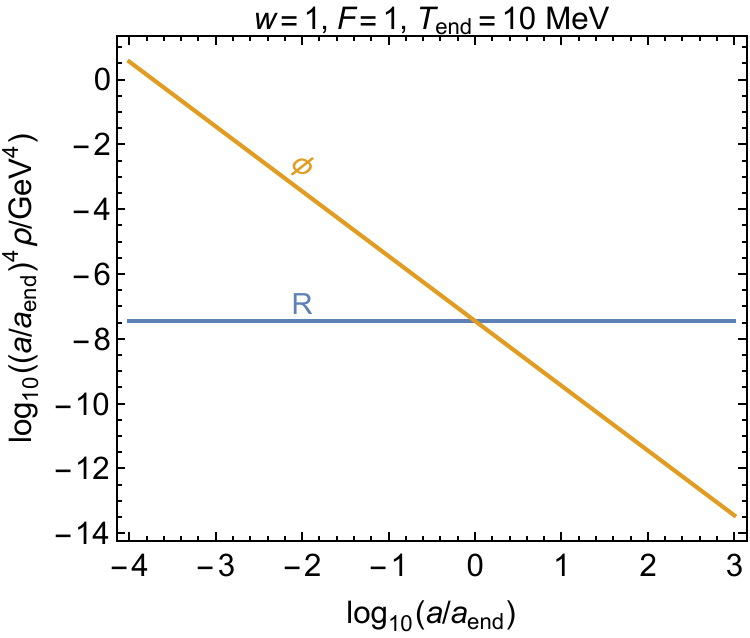}
	\caption{The solid yellow lines show the evolution of $\rho_\phi$ for $w=-1$, 0, 1 (upper left, upper right and lower panels, respectively), and the blue lines show the evolution of $\rho_{\rm R}$. Here $\aend \equiv a(T=\Tend)$ denotes the value of the scale factor $a$ when the $\rho_\phi$ dominance ends and the usual radiation-dominated phase begins. }
\label{fig:scaling}
\end{center}
\end{figure*}

A constant $\Gamma_\phi$ does not in all cases describe the evolution of the system well. For example, in the case where the Universe undergoes a (second) period of low-scale inflation, the system is better described by a step-function-type $\Gamma_\phi$ that gets a non-zero value at the end of a second inflationary phase. This can be realized for example if a phase transition ends the late inflationary phase, as in e.g. Ref.~\cite{Hambye:2018qjv}. Assuming that after the second inflationary period $\Gamma_\phi$ is larger than the Hubble rate, we can approximate that $\rho_\phi$ decays instantaneously to SM radiation. 
In this case
\be \label{rhoRstepf}
\frac{\rho_{\rm R}(a)}{\rho_{R}(\aend)} \simeq 
\begin{cases}
\left(\frac{a}{\aend}\right)^{-4} F^{-1} \,,\quad & a < \aend \,, \\
\left(\frac{a}{a_{\mathrm{end}}}\right)^{-4} \,, \quad &a \geq \aend \,.
\end{cases}
\ee
Also here $F$ equals the fraction of the co-moving SM radiation energy densities much before and much after the decay of $\rho_\phi$. 

In both of the above cases the evolution of $\rho_\phi$ and $\rho_{\rm R}$ can be completely described by three parameters: the $\rho_\phi$ equation of state parameter $w$, the
increase $F$ in the co-moving SM radiation energy density, and the temperature $T_{\rm end}$ of the SM radiation when the $\rho_\phi$ dominated period ends, 
which determines $\rho_{\rm R}(\aend)$ via Eq.~\eqref{rhoRT}. In the following, we will consider three benchmark cases numerically: 
\begin{enumerate}
\item $w=1$ and $\Gamma_\phi=0$, 
\item $w=0$ and a constant $\Gamma_\phi\ne 0$, 
\item $w=-1$ with instantaneous decay of $\rho_\phi$ at $a=\aend$. 
\end{enumerate}
The $w=1$ dominated epoch is know as kination~\cite{Ferreira:1997hj}. In that case, we have for simplicity taken $\Gamma_\phi=0$ because whenever $w>1/3$, the $\rho_\phi$ component will eventually become energetically subdominant to radiation regardless of the value of $\Gamma_\phi$. For previous works on a similar scenario, see e.g. Refs.~\cite{Kainulainen:2006wq,Figueroa:2016dsc,Dimopoulos:2018wfg}. The second case can be motivated by the usual particle decay, and the third one for example by a period of low-scale inflation, as discussed above. 

In Fig.~\ref{fig:scaling}, examples of the evolution of energy densities in these three cases are shown. In the case shown in the upper right panel the $\rho_\phi$ dominated period begins at $T=180\,{\rm GeV}$, whereas in the upper left panel it begins at $T=10\,{\rm MeV}$. In both cases the SM plasma temperature after the decay of $\rho_\phi$ is $T_{\rm end} = 10\,{\rm MeV}$. Notice that while in the case shown in the upper right panel the temperature decreases monotonically, in the upper left panel $T<10\,{\rm MeV}$ during the $\rho_\phi$ dominated period, and the decay of $\rho_\phi$ finally increases the temperature back to $10\,{\rm MeV}$. To present the maximal effect a non-standard expansion phase can have on DM production, both here and in the following the results are shown for $T_{\rm end}=10$~MeV, which is close to the BBN bound. However, the results can be easily generalized to higher values of $T_{\rm end}$. Note that even though all of the above cases can be motivated by scenarios considered in the literature, our analysis does not concentrate on any particular model besides the DM one, which we will discuss in the next subsection.

\subsection{Scalar singlet dark matter}

For DM we consider a simple model which, on top of the SM field content and the $\rho_\phi$ component, contains a real scalar singlet $S$ which is odd under a discrete $\mathbb{Z}_2$ symmetry, while all the other fields are even. This symmetry makes $S$ a viable DM candidate. The only interaction between $S$ and the SM sector is via the Higgs portal coupling $\lambda_{\rm HS}S^2|H|^2/2$, where $H$ corresponds to the SM Higgs doublet. The scalar potential containing only renormalizable terms is~\cite{McDonald:1993ex,Burgess:2000yq}
\be
\label{potential}
V = \mu_{\rm H}^2|H|^2 + \lambda_{\rm H}|H|^4 + \frac{\mu_{\rm S}^2}{2} S^2 + \frac{\lambda_{\rm S}}{4}S^4 + \frac{\lambda_{\rm HS}}{2} S^2 |H|^2 \,.
\ee
The condition $0<\mu_{\rm S}^2 + \lambda_{\rm HS}\,v^2/2\equiv M_{\rm S}^2$, where $v$ is the vacuum expectation value of the SM Higgs field, ensures that the $\mathbb{Z}_2$ is not broken spontaneously at the electroweak breaking vacuum.\footnote{For $\mu_{\rm S}^2<0$ the requirement that the electroweak breaking minimum is the global minimum of the potential gives a lower bound on the $S$ self-coupling~\cite{Vaskonen:2016yiu}, which at $M_{\rm S}^2\ll \lambda_{\rm HS} v^2/2$ is $\lambda_{\rm S} > 1.9\lambda_{\rm HS}^2$, leading to non-perturbative values of $\lambda_{\rm S}$ at large $\lambda_{\rm HS}$. This region is, however, not of interest because of the constraint on the Higgs boson invisible decay discussed below.} Then, the Higgs doublet mass parameter, $\mu_{\rm H}^2 = -\lambda_{\rm H} v^2$, and the Higgs doublet quartic coupling, $\lambda_{\rm H} = m_{\rm h}^2/(2v^2)$, are fixed by the observed values of the Higgs boson mass $m_{\rm h}\simeq 125$~GeV and the electroweak scale $v\simeq 246$~GeV. 

The total parameter space in our scenario is thus six-dimensional, consisting of three particle physics parameters,  $\lambda_{\rm S}$, $\lambda_{\rm HS}$ and $M_{\rm S}$, and three cosmological parameters, $w$, $F$ and $T_{\rm end}$. We assume that $\rho_\phi$ does not decay into $S$ but that its relic abundance is produced by freeze-out or freeze-in from the SM plasma. For recent works where a component similar to $\rho_\phi$ is allowed to decay also into scalar singlet DM, see Refs.~\cite{Drees:2017iod,Hardy:2018bph}.

The only collider signature of the SM extension under consideration arises from the invisible decay of the Higgs boson $h$. The corresponding branching ratio is constrained by the LHC searches to be BR$_\text{inv}\lesssim 0.24$ at the $2\sigma$ confidence level~\cite{Khachatryan:2016whc}. This places an upper bound on the decay width $\Gamma_{h\to SS}$. Using $\Gamma_{\rm h}\simeq 4.07$~MeV for the total Higgs decay width and
\be \label{eq:hSS}
\Gamma_{h\to SS} = \frac{\lambda_{\rm HS}^2 v^2}{32\pi m_{\rm h}} \sqrt{1-\frac{4M_{\rm S}^2}{m_{\rm h}^2}} \,
\ee 
for the Higgs decay into two $S$ particles, the bound on $\Gamma_{h\to SS}$ translates to a bound on the portal coupling, which for $M_{\rm S}\ll m_{\rm h}$ is $\lambda_{\rm HS} < 0.014$. For $M_{\rm S}>m_{\rm h}/2$, this kind of constraint obviously cannot be placed.

A further constraint on the model parameters arises from the direct DM searches. The effective spin-indepen\-dent cross section for 
elastic DM--nucleon scattering is given by
\begin{equation}
\sigma_\text{SI,\,eff}= \frac{\Omega_{\rm S}}{\Omega_{\rm DM}}\frac{\lambda_{\rm HS}^2\,\mu_N^2 m_N^2\,f_N^2}{4\pi\,M_{\rm S}^2\,m_{\rm h}^4}\,,
\end{equation}
where $\mu_N = m_N M_{\rm S} / (m_N+M_{\rm S})$ is the reduced mass of the DM-nucleon system with $m_N\simeq 0.946$~GeV the nucleon mass, $f_N\simeq 0.30$ is the form factor~\cite{Cline:2013gha},\footnote{For recent works related to the sigma terms, see Refs.~\cite{Farina:2009ez,Giedt:2009mr,Alarcon:2011zs,Ren:2012aj,Alarcon:2012nr,Ren:2014vea,Ling:2017jyz}.} and $\Omega_{\rm S}/\Omega_{\rm DM}$ is the fractional DM density. Currently, the most stringent constraints on $\sigma_\text{SI,\,eff}$ are provided by LUX~\cite{Akerib:2016vxi}, PandaX-II~\cite{Cui:2017nnn} and Xenon1T~\cite{Aprile:2018dbl}. The projected sensitivity of the next generation DM direct detection experiment DARWIN~\cite{Aalbers:2016jon} will also be shown in the following results.

\section{Dark matter abundance}
\label{sec:DMproduction}

In this section we discuss the production of DM in the early Universe during a non-standard expansion phase, considering first freeze-out and then freeze-in of DM. The strength of the portal coupling $\lambda_{\rm HS}$ determines whether the $S$ particles were in thermal equilibrium with the SM radiation in the early Universe. For the freeze-out mechanism the portal coupling has to be typically much larger than a threshold value  $\lambda_{\rm HS}\gg\lambda_{\rm HS}^\text{eq}$, whereas for freeze-in $\lambda_{\rm HS}\ll\lambda_{\rm HS}^\text{eq}$. The threshold value above which the DM sector enters into thermal equilibrium with the SM can be found by requiring that the SM particles do not populate the hidden sector so that they would start to annihilate back to the SM in large amounts~\cite{Petraki:2007gq,Enqvist:2014zqa,Kahlhoefer:2018xxo}. In the following, we will first consider freeze-out of DM by assuming that $\lambda_{\rm HS}\gg\lambda_{\rm HS}^\text{eq}$ always holds, and postpone a quantitative derivation of $\lambda_{\rm HS}^\text{eq}$ until Section \ref{sec:freezein}. In all cases in Section \ref{sec:freezeout}, however, the presented results have been found to be consistent with the thermalization condition.

Before discussing DM production mechanisms in mo\-re detail, we note that the observed DM abundance cannot be obtained for all expansion histories independently of the DM model parameters, assuming that the decay of $\rho_\phi$ does not produce $S$ particles. If $\rho_\phi$ dominates the energy density of the Universe when the co-moving $S$ number density freezes, then it can happen that the energy density in $S$ particles is always too small to comprise the observed DM abundance, unless the decay of $\rho_\phi$ brings $S$ back into thermal equilibrium in the case of freeze-out or re-triggers the freeze-in yield.\footnote{Notice that, independently on whether the $S$ abundance is determined by freeze-out or freeze-in, the energy density in $S$ particles is necessarily smaller than that of one relativistic degree of freedom in the radiation bath when its co-moving number density freezes.} However, for the benchmark scenarios discussed in Section~\ref{expansion}, this is never the case.

\subsection{Freeze-out}
\label{sec:freezeout}

We begin by studying the case where the DM has reached thermal equilibrium with the SM radiation. The DM abundance is then determined by the freeze-out mechanism and, in the absence of large DM self-interactions, the relevant interaction rate is that of DM annihilate into radiation bath particles, $\langle\sigma_\text{ann}v\rangle n_{\rm S}$. The contribution of the $hh$ final state at $s=4M_{\rm S}^2$ is given by~\cite{Alanne:2014bra}
\be
\begin{split}
&\langle\sigma_{hh}v\rangle = \frac{1}{32\pi s} \sqrt{1-\frac{4m_{\rm h}^2}{s}} \\
&\,\times\left|\lambda_{\rm HS}+\frac{3m_{\rm h}^2\lambda_{\rm HS}}{s-m_{\rm h}^2+ i m_{\rm h}\Gamma_{\rm h}}-\frac{4v^2\lambda_{\rm HS}^2}{s-2m_{\rm h}^2}\right|^2 \Bigg|_{s=4M_{\rm S}^2}
\end{split}
\ee
and all other SM final states can be taken into account by using the total decay width of virtual $h$, $\Gamma_{\rm h}(s)$, as~\cite{Cline:2012hg}
\be
\langle\sigma_{\rm SM}v\rangle = \frac{2\lambda_{\rm HS}^2 v^2 \Gamma_{\rm h}(s)}{\sqrt{s}\left[\left(s-m_{\rm h}^2\right)^2+m_{\rm h}^2 \Gamma_{\rm h}^2\right]}\Bigg|_{s=4M_{\rm S}^2} \,,
\ee
so that $\langle\sigma_\text{ann}v\rangle = \langle\sigma_{hh}v\rangle + \langle\sigma_{\rm SM}v\rangle$. 
The evolution of the $S$ number density is then described by the Boltzmann equation
\be\label{BEFO}
\frac{{\rm d}n_{\rm S}}{{\rm d}t}+3Hn_{\rm S} = -\langle\sigma_\text{ann} v\rangle\left[n_{\rm S}^2-(n_{\rm S}^\text{eq})^2\right] \,,
\ee
which we solve numerically. The time dependence of the Hubble parameter~\eqref{Hubble} and the equilibrium number density $n_{\rm S}^\text{eq}$ are obtained by solving the coupled Boltzmann equations for $\rho_\phi$ and $\rho_{\rm R}$ given by Eq.~\eqref{BE0}.

\begin{figure*}[t]
\begin{center}
\includegraphics[height=0.38\textwidth]{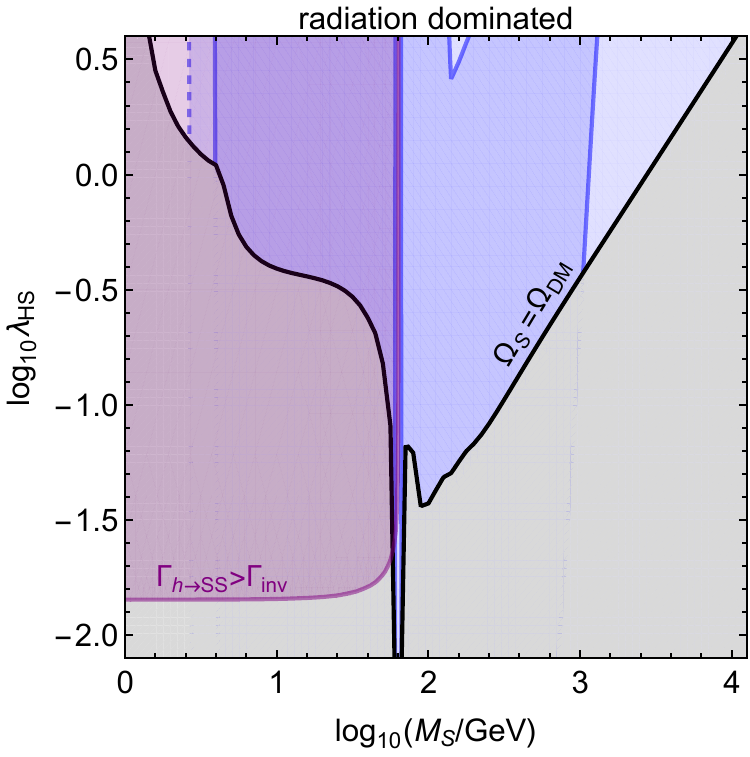} \hspace{6mm}
\includegraphics[height=0.38\textwidth]{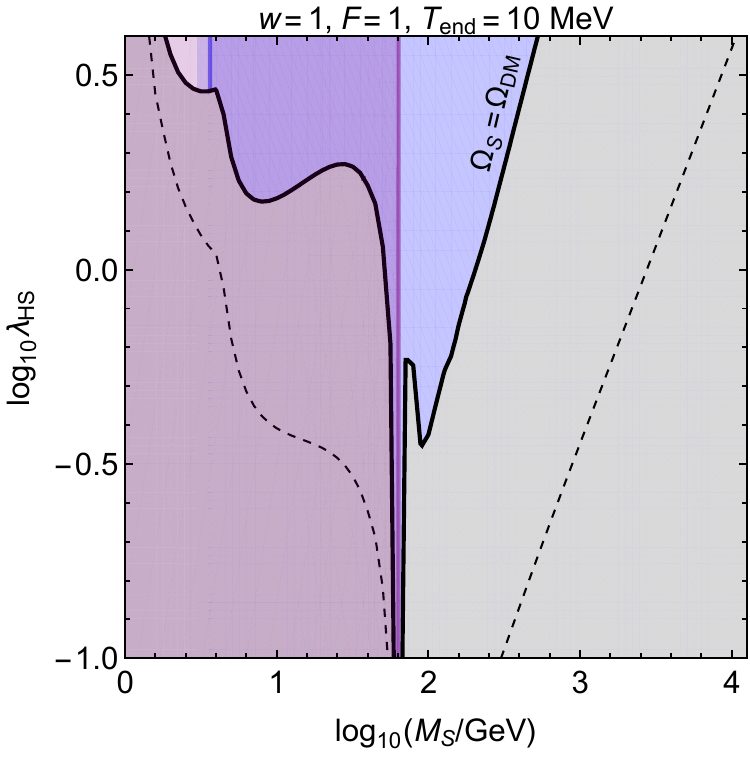} \\ \vspace{2mm}
\includegraphics[height=0.38\textwidth]{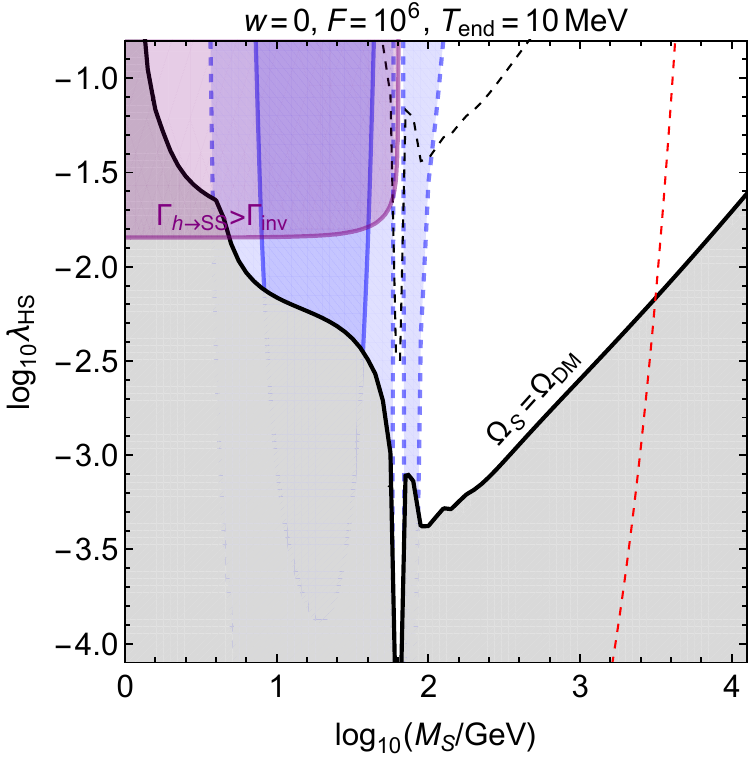} \hspace{6mm}
\includegraphics[height=0.38\textwidth]{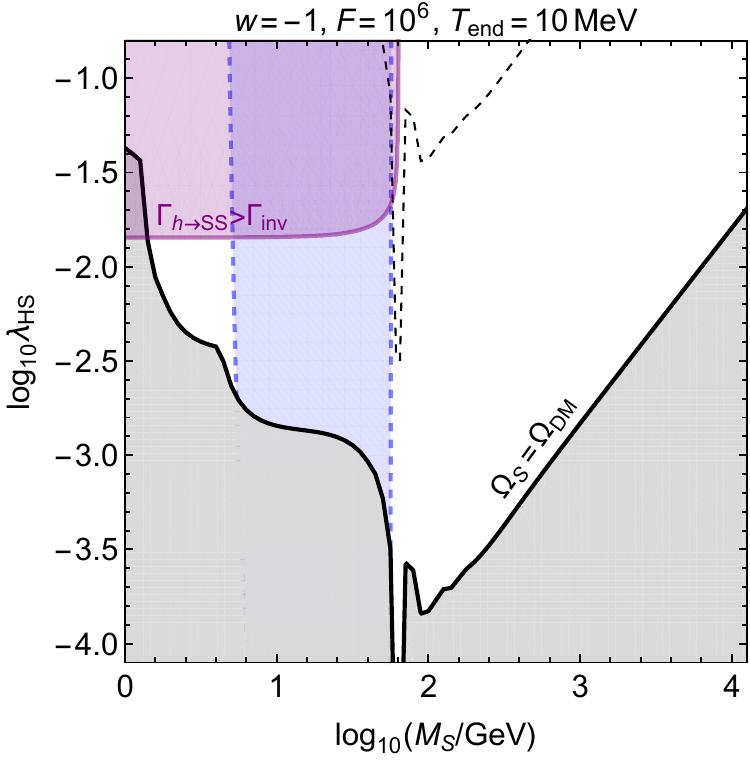}
\caption{In each panel, the solid black lines show the values of $\lambda_{\rm HS}$ and $M_{\rm S}$ for which the observed DM abundance for $S$ is obtained via freeze-out. The dark blue region is excluded by direct DM searches (LUX, PandaX-II and Xenon1T), the lighter blue region shows the expected sensitivity of the DARWIN experiment, and the purple region is excluded by the LHC constraint on the Higgs boson invisible decay. The dashed black line in each panel shows the standard radiation-dominated case. Left (right) from the red dashed line in the lower left panel the DM freeze-out happens during (before) the $\rho_\phi$ dominated phase. For all masses shown in the lower (upper) right panel the freeze-out happens before (during) the $\rho_\phi$ dominance.}
\label{fig:fo1}
\end{center}
\end{figure*}

By scanning the values of $\lambda_{\rm HS}$ and $M_{\rm S}$ for different background evolutions, we determine the value of the portal coupling for a given $S$ mass that gives the observed DM abundance, $\Omega_{\rm S}h^2 = \Omega_{\rm DM}h^2\simeq0.12$~\cite{Ade:2015xua}. This is shown by the black solid lines in Fig.~\ref{fig:fo1}. We see that a non-standard expansion phase can lead to a significant change in the DM abundance and therefore to observational ramifications. 

Compared to the standard radiation dominated case, two effects change the required value of the portal $\lambda_{\rm HS}$ for which the observed DM abundance is obtained: the moment when the co-moving DM number density free\-zes is shifted due to non-standard dependence of the Hubble parameter on the SM radiation temperature, and the DM energy density becomes effectively diluted due to decay of $\rho_\phi$. The effect of the former is to increase the required value of $\lambda_{\rm HS}$, whereas the latter decreases it. 

Consider first the case $F=1$ (upper right panel in Fig.~\ref{fig:fo1}). In this scenario no dilution due to $\rho_\phi$ decay arises, so the freeze-out temperature has to be the same as in the standard case in order to obtain the same final DM abundance. Then, if the DM freeze-out happens when $\rho_\phi$ dominates the energy density of the Universe, the interaction rate that keeps DM in thermal equilibrium has to be higher than in the standard case, because the value of the Hubble parameter at that temperature is higher. This implies that $\lambda_{\rm HS}$ has to be larger than in the standard radiation dominated case. Depending how the ratio $\rho_{\rm R}/\rho_\phi$ evolves as a function of $T$, this increase of $\lambda_{\rm HS}$ is different for different masses. The upper right panel of Fig.~\ref{fig:fo1} shows an example of such a scenario. In that case the ratio $\rho_{\rm R}/\rho_\phi$ increases as a function of $T$ so for large masses the separation between the black solid and dashed lines is larger than for small masses.

Moreover, for $F>1$ the freeze-out temperature has to be higher, because the decay of $\rho_\phi$ decreases the relative DM energy density compared to $\rho_{\rm R}$. Thus, the same final abundance is obtained by decreasing the value of $\lambda_{\rm HS}$ so that the $S$ particles undergo freeze-out earlier, which leads to the required enhancement in the co-moving $S$ number density. In the lower right panel of Fig.~\ref{fig:fo1}, the DM freeze-out happens for all masses before the $\rho_\phi$ dominated phase begins, so the effect from the non-standard temperature dependence of the Hubble parameter is absent. In contrast to this, left from the red dashed line in the lower left panel the freeze-out happens during the $\rho_\phi$ dominance, but the effect from dilution due to the $\rho_\phi$ decay is still the dominant one.

In the usual radiation-dominated case only DM mas\-ses close to $m_{\rm h}/2$ and above $\mathcal{O}(1)\,{\rm TeV}$ are still allowed by observations~\cite{Cline:2013gha,No:2013wsa,Alanne:2014bra,Robens:2015gla,Feng:2014vea,Han:2015hda,Athron:2017kgt}. This can be seen in the upper left panel of Fig.~\ref{fig:fo1}, where the dark blue regions are excluded by direct DM searches (LUX, PandaX-II and Xenon1T), the lighter blue regions show the expected sensitivity of the DARWIN experiment, and the purple regions are excluded by the LHC constraint on the Higgs boson invisible decay. This conclusion changes in the case DM was produced during a non-standard expansion phase. In particular, we see that in the cases shown in the lower panels, large parts of the parameter space become available. In the case shown in the upper right panel, however, the required values of $\lambda_{\rm HS}$ are larger than in the usual radiation-dominated case, which renders that scenario largely inconsistent with observations. While these conclusions may change in models which go beyond the benchmark scenarios discussed in Section~\ref{expansion}, they demonstrate the fact that a non-standard expansion history can change the requirements for producing the observed DM abundance in interesting and yet testable ways.

Finally, to conclude the discussion about the DM production via freeze-out, we remark that while in the standard radiation-dominated case the $S$ self-coupling $\lambda_{\rm S}$ has to take non-perturbative values for the $S$ number changing self-interactions (such as $SSSS\to SS$) to determine the $S$ freeze-out instead of the $SS$ annihilations to SM particles~\cite{Bernal:2015xba}, these processes can be relevant in non-standard cases. If the dilution due to the 
decay of $\rho_\phi$ to $\rho_{\rm R}$ after the $S$ freeze-out was sufficiently strong, the $SSSS\to SS$ processes become relevant even for $\lambda_{\rm S} \lesssim 1$. Therefore, taking the detailed effect of DM self-interactions into account can be important for the determination of the final DM abundance, reminiscent to the Strongly Interacting Massive Particle (SIMP) or cannibal DM scenarios~\cite{Bernal:2018ins,Bernal:2015xba,Dolgov:1980uu,Carlson:1992fn,Hochberg:2014dra,Bernal:2015bla,Bernal:2015lbl,Bernal:2015ova,Pappadopulo:2016pkp,Heikinheimo:2016yds,Farina:2016llk,Chu:2016pew,Dey:2016qgf,Bernal:2017mqb,Choi:2017mkk,Heikinheimo:2017ofk,Ho:2017fte,Dolgov:2017ujf,Garcia-Cely:2017qpx,Hansen:2017rxr,Chu:2017msm,Duch:2017khv,Chauhan:2017eck,Herms:2018ajr,Heikinheimo:2018esa,Hochberg:2018vdo,Hochberg:2018rjs}. A benchmark example of such a scenario was recently studied in Ref.~\cite{Bernal:2018ins}, but in this paper we do not consider the detailed effect of the $SSSS\to SS$ process. Instead, we have assumed that $\lambda_{\rm S}$ is always small enough not to affect DM production to highlight the effects of non-standard expansion history.

\subsection{Freeze-in}
\label{sec:freezein}

Next, we turn to the case where the $S$ particles interact so feebly with the SM radiation that they never entered into thermal equilibrium with it, and the relevant production mechanism is freeze-in. Assuming that the $S$ abundance is always negligible compared to its equilibrium abundance, the Boltzmann equation describing the $S$ production from annihilations of the SM particles and the Higgs boson decay is 
\be\label{BEFI}
\begin{split}
\frac{\td n_{\rm S}}{\td t}+3\,H\,n_{\rm S} =& \sum_x \langle \sigma_{x\bar{x}\to SS} v\rangle (n_x^{\rm eq})^2 \\
&+ C\, m_{\rm h} \,\Gamma_{h\to SS} \int \frac{\td^3 p_{\rm h}}{E_{\rm h} (2\pi)^3} f_{\rm h}^{\rm eq} \,,
\end{split}
\ee
where the sum runs over all SM particles. The $h\to SS$ decay width is given by Eq.~\eqref{eq:hSS}, and annihilation cross sections are\footnote{The factors $g_x^{-2}$ arise from averaging over the initial states.}
\be
\begin{aligned}
\sigma_{VV\to SS} =& g_V^{-2} \frac{\lambda_{\rm HS}^2}{32\pi s} \sqrt{\frac{s-4M_{\rm S}^2}{s-4M_V^2}} \frac{\left(s^2-4s M_V^2+12M_V^4\right)}{(s-m_{\rm h}^2)^2+m_{\rm h}^2 \Gamma_h^2} \,,\\
\sigma_{f\bar{f}\to SS} =& g_f^{-2} \frac{\lambda_{\rm HS}^2 n_c}{16\pi s} \frac{\sqrt{(s-4M_f^2)(s-4M_{\rm S}^2)} M_f^2}{(s-m_{\rm h}^2)^2+m_{\rm h}^2 \Gamma_h^2} \,,\\
\sigma_{hh\to SS} =& \frac{1}{32\pi s} \sqrt{\frac{s-4M_{\rm S}^2}{s-4m_{\rm h}^2}} \\ &\times\left|\lambda_{\rm HS}+\frac{3m_{\rm h}^2\lambda_{\rm HS}}{s-m_{\rm h}^2+ i m_{\rm h}\Gamma_{\rm h}}-\frac{4v^2\lambda_{\rm HS}^2}{s-2m_{\rm h}^2}\right|^2 \,,
\end{aligned}
\ee
where $V=W^\pm,\,Z$ and $f$ ($\bar{f}$) denote the SM (anti-) fermions, and $n_c=3$ for quarks and $n_c=1$ for leptons. The dominant production channels are $W^+W^-\to SS$ for $M_{\rm S}>m_{\rm h}/2$ and $h\to SS$ for $M_{\rm S}<m_{\rm h}/2$.

The contribution of Higgs boson decays is partly included in the on-shell part of the $x\bar{x}\to SS$ annihilation processes (see e.g. Ref.~\cite{Frigerio:2011in}), and thus subtracted from the decay term in Eq.~\eqref{BEFI} by multiplying it by 
\be
C = 1-\sum_x {\rm BR}(h\to x\bar{x})\,.
\ee
Here the sum includes SM particles with mass $m_x < m_{\rm h}/2$, i.e. it does not include virtual final states. The dominant channels are then to $b\bar{b}$, $\tau^-\tau^+$ and $c\bar{c}$, with branching ratios ${\rm BR}(h\to b\bar{b})\simeq 0.561$, ${\rm BR}(h\to \tau^-\tau^+)\simeq 0.0615$ and ${\rm BR}(h\to c\bar{c})\simeq 0.0283$, so $C\simeq 0.349$. 

\begin{figure}[t]
\begin{center}
\includegraphics[height=0.38\textwidth]{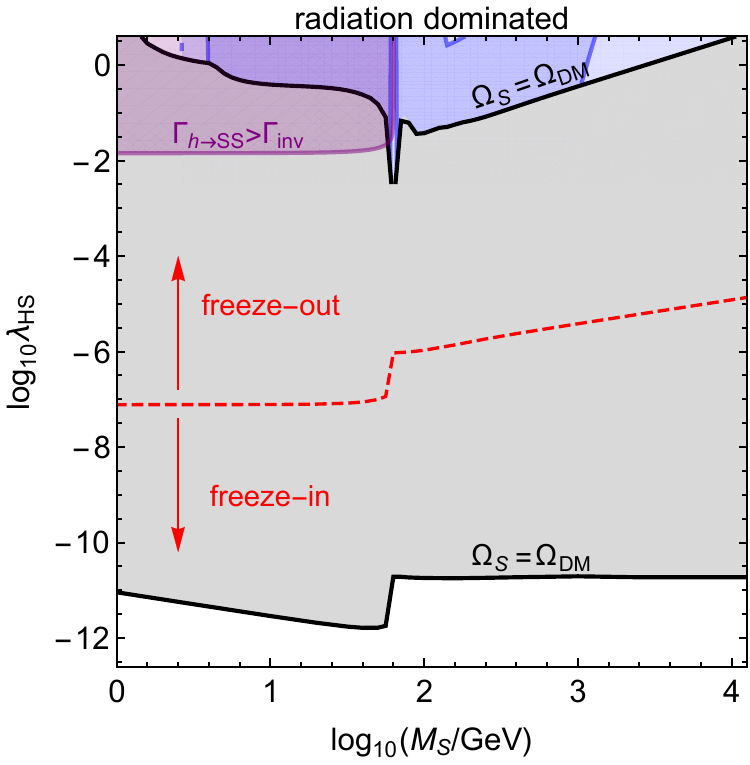} 
\caption{The red dashed line shows the thermalization bound for $\lambda_{\rm HS}$. The colored regions are the same as in Fig.~\ref{fig:fo1}, and the upper and lower solid black lines show the values of $\lambda_{\rm HS}$ which give the observed DM abundance via freeze-out and freeze-in, respectively.}
\label{fig:thermalisation}
\end{center}
\end{figure}

As discussed in the beginning of Section~\ref{sec:DMproduction}, the threshold value above which the DM sector enters into thermal equilibrium with the SM can be found by requiring that the SM particles do not populate the hidden sector so that they would start to annihilate back to the SM in large amounts. As the criterion for this, we require that at all times
\be
\sum_x \frac{\langle \sigma_{x\bar{x}\to SS} v\rangle n_x^{\rm eq}}{H} < 1\,.
\ee
The threshold value $\lambda_{\rm HS}^{\rm eq}$ can then be found as the smallest $\lambda_{\rm HS}$ for which the above ratio reaches unity. For example, in the usual radiation-dominated case this threshold is shown by the red dashed line in Fig. \ref{fig:thermalisation}, which is in accord with the earlier estimate in the literature~\cite{Petraki:2007gq,Enqvist:2014zqa,Kahlhoefer:2018xxo}. However, because in non-standard cases $H$ takes a larger value than in the usual radiation-domination, also $\lambda_{\rm HS}$ can be larger than usual without the two sectors thermalizing with each other. In the following, we check that $\lambda_{\rm HS}<\lambda_{\rm HS}^{\rm eq}$ in all cases.

\begin{figure*}[t]
\begin{center}
\includegraphics[height=0.32\textwidth]{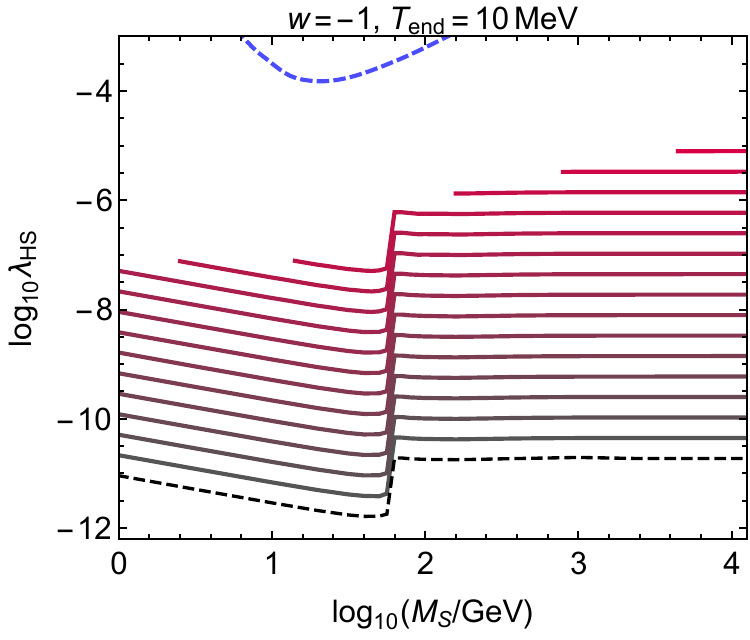} \hspace{6mm}
\includegraphics[height=0.32\textwidth]{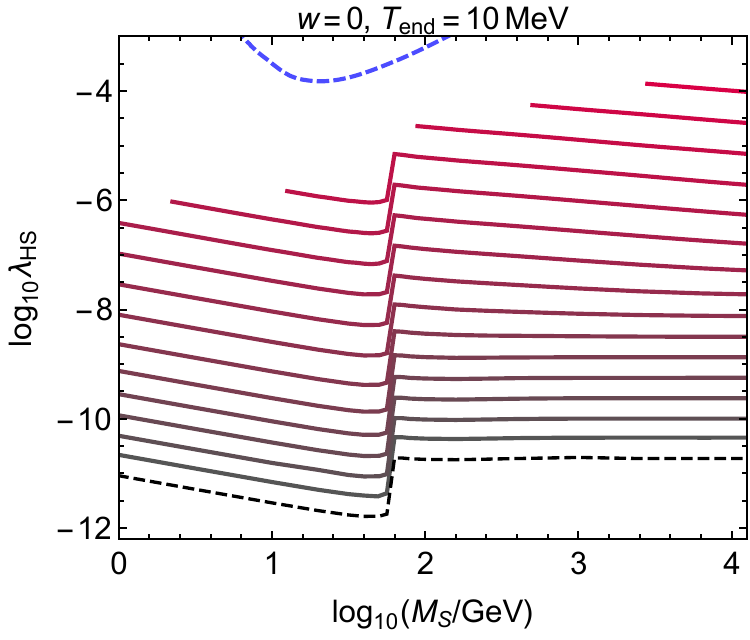} \\ \vspace{2mm}
\includegraphics[height=0.32\textwidth]{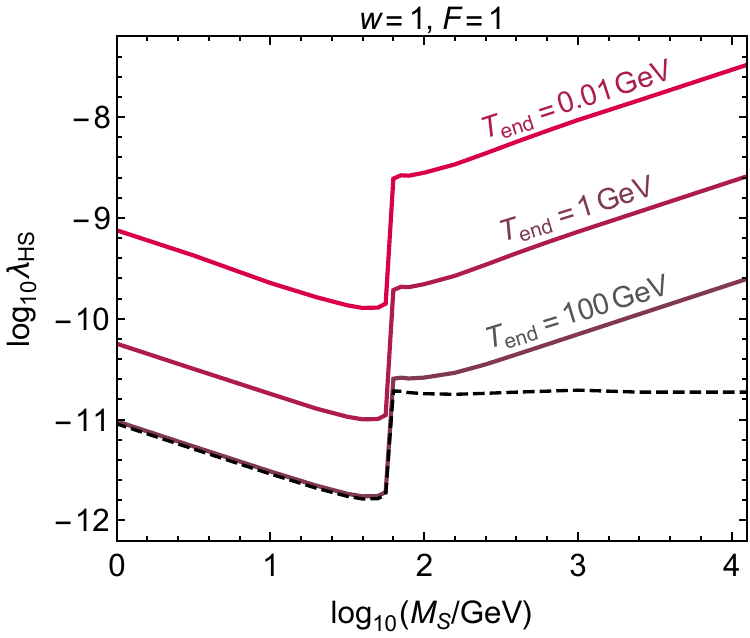}
\caption{The curves show the value of $\lambda_{\rm HS}$ for which the observed DM abundance in $S$ particles is obtained via freeze-in for different expansion histories with $w=-1$, 0, 1. In the top panels the solid lines correspond to $\log_{10} F = 1$, 2\dots15, from bottom to top.
For large values of $F$ and  small $M_{\rm S}$, the freeze-in picture is not consistent (thermalization with the SM is reached), which introduces a cut-off to the corresponding curves. The black dashed curve in the bottom of each panel shows the result in the standard radiation dominated case.
The region above the blue curve is expected to be probed by DARWIN, if all DM is in $S$ particles.}
\label{fig:fi}
\end{center}
\end{figure*}

Defining $Y_{\rm S} = a^3 n_{\rm S}$ allows us to solve Eq.~\eqref{BEFI}. Then, the $S$ abundance today is $\Omega_{\rm S} = M_{\rm S} Y_{\rm S}(a=1)/\rho_c$, where $\rho_c$ is the critical density and the co-moving $S$ number density today is given by
\be
\begin{split}
Y_{\rm S}(a=1) &= \int_0^1 \td a\,\frac{a^2}{H} \bigg[\frac{T}{32\pi^4} \int_{4M_{\rm S}^2}^\infty \td s \sum_x c_x\, g_{\rm x}^2\\
&\times\sigma_{x\bar{x}\to SS}(s) (s-4m_x^2)\,\sqrt{s} \,K_1(\sqrt{s}/T) \\
& + \frac{g_{\rm h}\, m_{\rm h}^2\, T}{2\pi^2} \,C\,\Gamma_{h\to SS}\, K_1(m_{\rm h}/T) \bigg] \,.
\end{split}
\ee
Here $c_x=1/2$ if the initial state particles are identical and $c_x=1$ otherwise, and $g_x$ is the number of degrees of freedom for particle species $x$.\footnote{Particles and antiparticles are treated separately here. Thus, for gauge bosons $g_V = 3$, for leptons $g_l = 2$, for quarks $g_q = 6$ and for the Higgs boson $g_h=1$.} The Hubble parameter~\eqref{Hubble} and the radiation bath temperature $T$ are again obtained as a function of the scale factor $a$ by solving the coupled Boltzmann equations~\eqref{BE0}. 

The results are shown in Fig.~\ref{fig:fi}. As in the freeze-out case, also here the effects that shift the required value of $\lambda_{\rm HS}$ are the non-standard dependence of the Hubble parameter on the SM radiation temperature and the effective dilution of the DM energy density due to the decay of $\rho_\phi$. However, both of these effects now increase the required value of $\lambda_{\rm HS}$. If $F=1$ (lower panel of Fig.~\ref{fig:fi}), the DM production rate has to be higher than in the standard case because the value of the Hubble parameter at the temperature when the production ends (that is determined by the masses of the decaying/annihilating particles and $S$) is higher. Larger values of $F$ then imply that the relative DM energy density gets smaller due to the decay of $\rho_\phi$ to $\rho_{\rm R}$, so the DM production rate has to be even higher in order to obtain the same final abundance. This effect can be seen in the upper panels of Fig.~\ref{fig:fi}, where $\log_{10}F=1$, 2\dots 15, from bottom to top.

We find that for the entire mass regime studied in this paper, 1~GeV $\leq M_{\rm S}\leq 10$~TeV, the portal coupling which does not thermalize $S$ with the SM sector but allows it to constitute all of the observed DM abundance is always below the expected sensitivity of DARWIN (blue line in Fig.~\ref{fig:fi}). Therefore, we conclude that in the singlet scalar model even very extreme scenarios, where the expansion rate of the Universe exceeds the one in usual radiation-domination by many orders of magnitude, yield no observable consequences for next generation direct detection experiments in our benchmark scenarios where DM was produced by freeze-in.

Finally, we make again a remark on the effect of $S$ self-interactions. Also in the freeze-in scenario the effect of $SSSS\to SS$ self-annihilation can have important consequences on the final DM abundance after the initial yield from the SM sector has shut off. However, as noticed in Refs.~\cite{Bernal:2018ins,Bernal:2015xba,Bernal:2015ova,Heikinheimo:2016yds,Heikinheimo:2018esa}, the effect of this process is to generically increase the final DM abundance, which means that in the case where the number-changing self-annihilations play a role, a smaller value of $\lambda_{\rm HS}$ than in scenarios where self-interactions are absent is required to obtain the observed DM abundance. As the largest possible values of $\lambda_{\rm HS}$ are below the ones that can be expected to be detected by the next generation experiments, we have again chosen to restrict our analysis to values of $\lambda_{\rm S}$ which are small enough not to affect the DM yield.

\section{Conclusions}
\label{sec:conclusions}

Despite the large amount of searches over the past de\-cades, DM has not been found. A simple reason for this might be that the cosmological history was non-standard at early times, affecting also DM genesis. In this paper we have considered production of DM in such a scenario, studying both the freeze-out and freeze-in mechanisms in a model where the DM consists of scalar singlet particles.

Assuming that the DM number-changing interactions can be neglected, we showed in three benchmark scenarios that in the case of non-standard expansion history, two effects change the required value of the portal $\lambda_{\rm HS}$ for which the observed DM abundance is obtained: the moment when the co-moving DM number density freezes is shifted due to non-standard dependence of the Hubble parameter on the SM radiation temperature and, assuming that the dominant energy density component decayed solely to SM radiation after DM production, the DM energy density becomes effectively diluted. The effect of the former is to increase the required $\lambda_{\rm HS}$ in both freeze-out and freeze-in cases, whereas the latter in the freeze-out case decreases $\lambda_{\rm HS}$, and in the freeze-in case increases it. 

These findings, as well as the detailed changes to the allowed part of the parameter space together with prospects for future observations, are shown in Figs.~\ref{fig:fo1} and~\ref{fig:fi}. While these conclusions may change in models which go beyond the benchmark scenarios discussed in this paper, the results demonstrate the fact that a non-standard expansion history can change significantly the requirements for producing the observed DM abundance. 
For example, we find that in the freeze-out case the direct detection constraints in the singlet scalar DM model can be avoided if the early Universe was dominated by a matter-like component for a relatively short period of time before BBN. However, our results show that the parameter space relevant for freeze-in in the singlet scalar model is out of reach of the next generation direct detection experiments even in very extreme scenarios. 

In the future, it would be interesting to see what are the detailed consequences of non-standard expansion history also for other models where the hidden DM sector has more structure or where the DM is not coupled to the SM via the Higgs portal but by some other means.

\section*{Acknowledgments}
We thank X. Chu, C.S. Fong, T. Hambye, M. Heikinheimo, and L. Marzola for discussions. C.C. is supported by the Funda\c{c}\~{a}o para a Ci\^{e}ncia e Tecnologia (FCT) grant PD/BD/114453/2016, T.T. by the U.K. Science and Technology Facilities Council grant \linebreak ST/J001546/1, V.V. by the Estonian Research Council Grant No. IUT23-6 and ERDF Centre of Excellence Project No. TK133, and N.B. partially by Spanish MINECO under Grant FPA2017-84543-P. This project has also received funding from the European Union’s Horizon 2020 research and innovation programme under the Marie Skłodowska-Curie grant agreements \linebreak 674896 and 690575; and from Universidad Antonio \linebreak Nariño grants 2017239 and 2018204. N.B. and C.C. acknowledge the hospitality of the IFIC and the University of Helsinki, respectively.

\bibliography{DMcosmo}

\end{document}